\documentclass[sigconf]{acmart}

\usepackage{bbding}
\usepackage{multirow}
\usepackage{amsmath}
\usepackage{xcolor}

\usepackage{lipsum}
\AtBeginDocument{%
  \providecommand\BibTeX{{%
    \normalfont B\kern-0.5em{\scshape i\kern-0.25em b}\kern-0.8em\TeX}}}

\copyrightyear{2024}
\acmYear{2024}
\setcopyright{acmlicensed}\acmConference[In2Writing '24]{The Third Workshop on Intelligent and Interactive Writing Assistants}{May 11, 2024}{Honolulu, HI, USA}
\acmBooktitle{The Third Workshop on Intelligent and Interactive Writing Assistants (In2Writing '24), May 11, 2024, Honolulu, HI, USA}
\acmDOI{10.1145/3690712.3690723}
\acmISBN{979-8-4007-1031-5/24/05}

\begin{document}

\title[Sense of Ownership and Reasoning]{LLMs as Writing Assistants: Exploring Perspectives on Sense of Ownership and Reasoning}

\author{Azmine Toushik Wasi}
\authornote{Contact author.}
\orcid{0000-0001-9509-5804}
\affiliation{%
  \institution{Shahjalal University of Science and Technology}
  \city{Sylhet}
  \country{Bangladesh}
  }
  \email{azminetoushik.wasi@gmail.com}

\author{Mst Rafia Islam}
\orcid{/0009-0000-2561-3418}
\affiliation{%
  \institution{Independent University}
  \city{Dhaka}
  \country{Bangladesh}
  }
\email{rafiabarsha21@gmail.com}

\author{Raima Islam}
\orcid{0009-0007-4744-9015}
\affiliation{%
  \institution{BRAC University}
  \city{Dhaka}
  \country{Bangladesh}
}
\email{raima.islam@g.bracu.ac.bd}






\renewcommand{\shortauthors}{Wasi et al.}

\begin{abstract}
Sense of ownership in writing confines our investment of thoughts, time, and contribution, leading to attachment to the output. However, using writing assistants introduces a mental dilemma, as some content isn't directly our creation. For instance, we tend to credit Large Language Models (LLMs) more in creative tasks, even though all tasks are equal for them. Additionally, while we may not claim complete ownership of LLM-generated content, we freely claim authorship. We conduct a short survey to examine these issues and understand underlying cognitive processes in order to gain a better knowledge of human-centered aspects in writing and improve writing aid systems.
\end{abstract}





\maketitle
\section{Introduction}
Ownership and a sense of ownership are fundamental aspects of human cognition and behavior, ingrained from early childhood through a naive theory that includes ontological commitments and causal-explanatory reasoning  \cite{NANCEKIVELL2019102}. It implies that an individual has authority over a particular object, content, asset, or right. On the contrary, a widespread definition of \textit{sense of ownership} is, it is the feeling of responsibility and pride associated with being an owner or part of a team, project, or organisation. A person's psychological and emotional attachment to what they own or have control over is known as their sense of ownership \cite{doi:10.1037/1089-2680.7.1.84}. It goes beyond the legal or formal aspects to include the psychological and subjective experiences of \textit{ownership}. When it comes to writing or creative work, a person's sense of ownership is always linked to the degree of originality or effort needed to complete the writing task \cite{BarrattPugh2020}.

\begin{figure}[t] 
\centering {\includegraphics[scale=0.18]{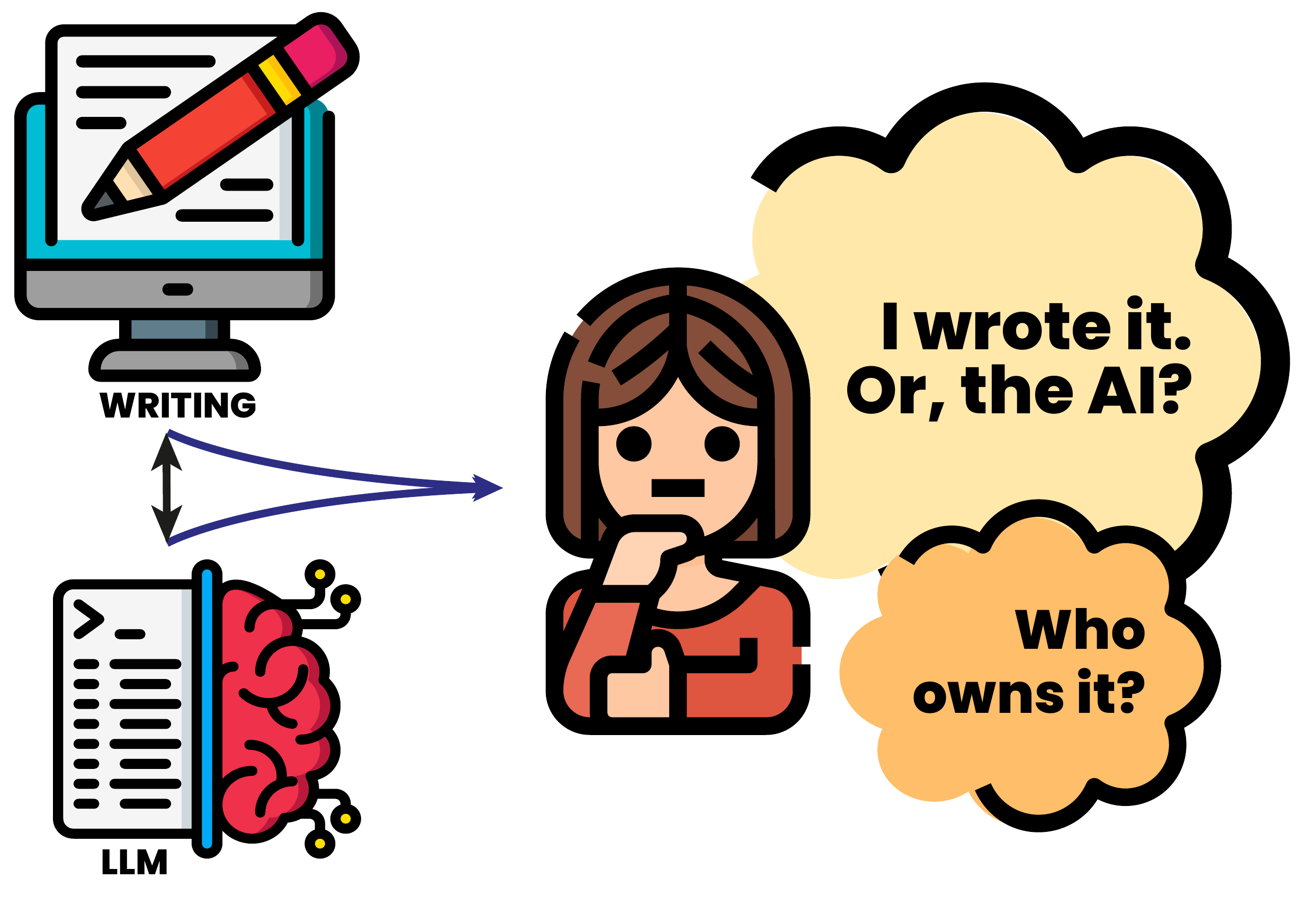}}
\caption{Mental dilemmas while using LLMs as Writing Assistants: \textit{Who wrote it? Who owns it?}}
\label{fig:WHAT}
\end{figure}

LLMs serve both creative and non-creative tasks, prompting the question: do our feelings of ownership align similarly across both creative and non-creative tasks when working with LLMs? As humans, we often feel more attached to creative tasks, attributing a greater sense of ownership to them \cite{Cassidy2013}. Another focus is investigating the link between a sense of ownership and authorship in LLM-based writing assistants, which is crucial, especially given the prevalent issue of plagiarism among humans \cite{park2017other,milano2023large-4}. Also, different social dynamics are involved here too, as indicated by \citet{10.1145/3544548.3580782}. Understanding how individuals perceive ownership and authorship in the context of LLM assistance can help address concerns about originality and attribution in content creation.

So, the first question is, \textbf{\textit{Does sense of ownership depend on the type of content while using LLMs as writing assistants?}}
When using LLMs for writing assistance, our sense of ownership varies depending on the type of content we're creating. Our level of involvement in idea generation and story formation influences our ownership feelings. We feel more ownership when actively shaping the content, but less so when relying heavily on LLM-generated material. Additionally, if the LLM autonomously completes a task, we tend to credit the model more, reducing our sense of ownership. Conversely, in routine, non-creative tasks, we feel a greater sense of contribution when collaborating with the LLM, though the level of work done by LLM is almost the same.

The second question is, \textbf{\textit{How does the use of LLM-based writing assistants affect perceptions of ownership and authorship?}}
Understanding our relationship with ownership also becomes complex when employing LLMs. Although we technically become the authors of content with their assistance, we may not feel as emotionally connected to it as if we created it independently. In reality, we may perceive ourselves as the authors of AI-generated content while also recognizing that we don't have full ownership over it \cite{10.1145/3637875,milano2023large-4}.
This discrepancy between authorship and ownership can create a mental dilemma while using LLM-generated content.

Understanding the complexities in our perception of ownership and authorship highlights the importance of addressing them when developing writing assistants. We conducted a brief survey with 35 participants to gain insight into and explore their perspectives to try to understand these complexities better. By understanding these, we hope to create more effective tools that align with users' needs and enhance their writing experiences. Additionally, navigating these complexities can help us use LLMs more efficiently in real-life scenarios, ensuring they serve as valuable aids rather than obstacles in the creative process.

\section{Cognition, Ownership, and the Rise of Language Models}
\subsection{Sense of Ownership in Writing.}
Ownership in writing transcends the physical act of possessing pen and paper; it's a profound bond between thoughts and words. Crafting sentences and paragraphs involves investing in ourselves, nurturing a sense of ownership over our ideas. It's more than claiming credit; it's cultivating a relationship with our creation, akin to tending a cherished plant. Each pen stroke embodies our identity, manifesting thoughts and emotions onto the page. This ownership empowers us to share our stories, knowing they carry our essence. In content writing, this sense of ownership arises from a complex interplay of cognitive and emotional experiences \cite{ElizabethKill}, weaving personal commitment and connection into every word, whether it's an assignment, a birthday wish, or an article. 

\subsection{Sense of Ownership in LLM Assisted Writing.} LLMs are becoming increasingly popular in both education and business because of their exceptional performance in a wide range of applications \cite{guo2023evaluating}.
LLMs can be used effectively in different writing tasks like creative stories and essays \cite{huang2023inspo,wordcraft,10.1145/3491102.3501819}, academic writing \cite{Bekker2024ww,Jarrah2023}, and legal write-ups \cite{nay2023large}. When utilizing LLMs like ChatGPT, Gemini or Microsoft Copilot, the sense of ownership in writing may evolve, as the process becomes collaborative, blending human input with machine-generated content. Additionally, LLMs can facilitate the exploration of diverse perspectives and styles, altering the traditional notions of authorship and creativity in content creation. 

\section{Exploring Problems and Perspectives}
We conducted a brief survey with 35 participants to evaluate their sense of ownership while using writing assistants, and the questions that arose above.  The demographic profile of the participants are available in Appendix  \ref{sec: demographics}.

\subsection{Survey Questions} \label{sec: sur-q}
The questions asked in the survey are added below:
\begin{itemize}
    \item General Writing: If you've utilized ChatGPT to compose an assignment, which is non-creative writing, do you believe you've made contributions to the content?
    \item Creative Writing: If you've utilized ChatGPT to compose a story or poem which is a creative task, do you believe you've made contributions to the content?
    \item Self-Declare as Authors: Will you submit any LLM-generated content somewhere with your name in it? (like assignments, newspaper article or anything)
    \item Sense of Ownership: Do you think you own the content ChatGPT made for you?
    \item Sense of Ownership (after reminding them about prompting): When considering the content generated by ChatGPT in response to specific prompts, there's an intricate interplay between the prompt itself and the resulting text. In this dynamic, prompts serve as catalysts for the AI model's generation, shaping the direction and nature of the response. However, given that the content is prompted and guided by human input, do you think you own the content ChatGPT made for you?
    \item Sense of Ownership (in modified LLM content): While the initial content was generated by ChatGPT, the modifications reflect a human intervention, shaping and refining the text to better fit the context or convey a specific message. In this altered state, as you have modified the original ChatGPT response; does that mean you own it now?
\end{itemize}

\subsection{Type of Content and Ownership} 
\begin{figure}[t] 
\centering {\includegraphics[scale=0.4]{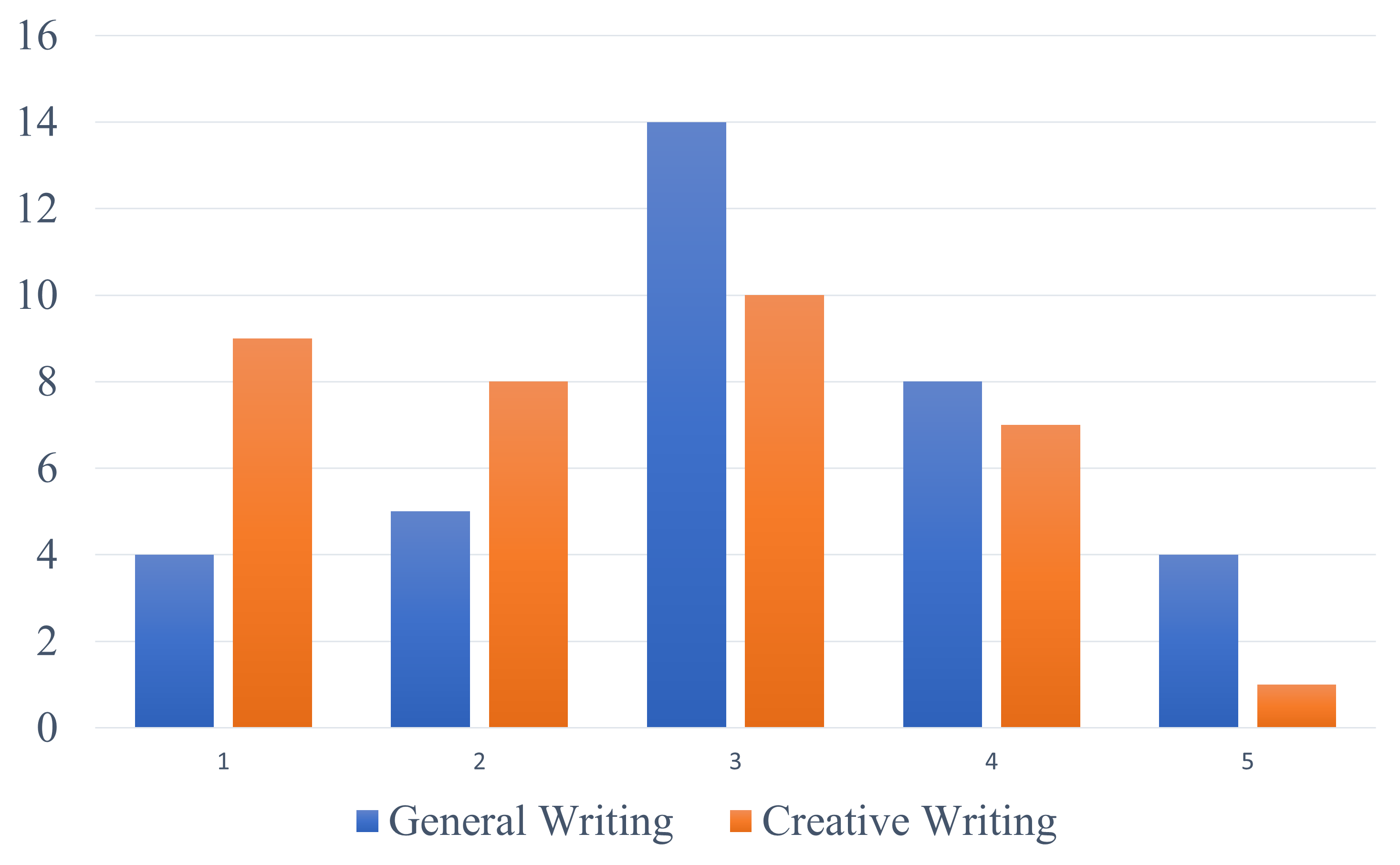}}
\caption{Survey Statistics: Participants answering how much contribution (1 being very low, and 5 being very high) they have in different types of LLM-assisted Writings}
\label{fig:9_10}
\end{figure}

Our survey indicates that individuals feel a stronger sense of contribution and ownership in non-creative content, such as STEM assignments and formal letters, where they perceive less involvement from the AI writing assistant. Conversely, in creative writing like poems, stories, and birthday wishes, participants tend to feel less ownership as they perceive the AI's contribution to be more significant. In Figure \ref{fig:9_10}, non-creative writing exhibits a generally normal distribution, slightly left-skewed. However, in creative writing, there is a notable bias towards attributing less human contribution, with individuals giving LLMs more credit. In creative writing, people often give one individual more credit than in non-creative writing historically \cite{articleetgyewy}.

In creative tasks, individuals appear to regard LLMs more similar to human contributors, as ownership and authorship are typically associated with originality, contribution, and accountability \cite{author-guide}. While LLMs can facilitate some levels of originality \cite{franceschelli2023creativity}, their accountability is questioned currently \cite{huang2023citation}. In contrast, in non-creative tasks, where originality and contribution are less critical, this may be less of a concern, and humans also tend to give them less credit too—an interesting observation.

\subsection{Authorship and Ownership} 
\begin{figure}[t] 
\centering {\includegraphics[scale=0.2]{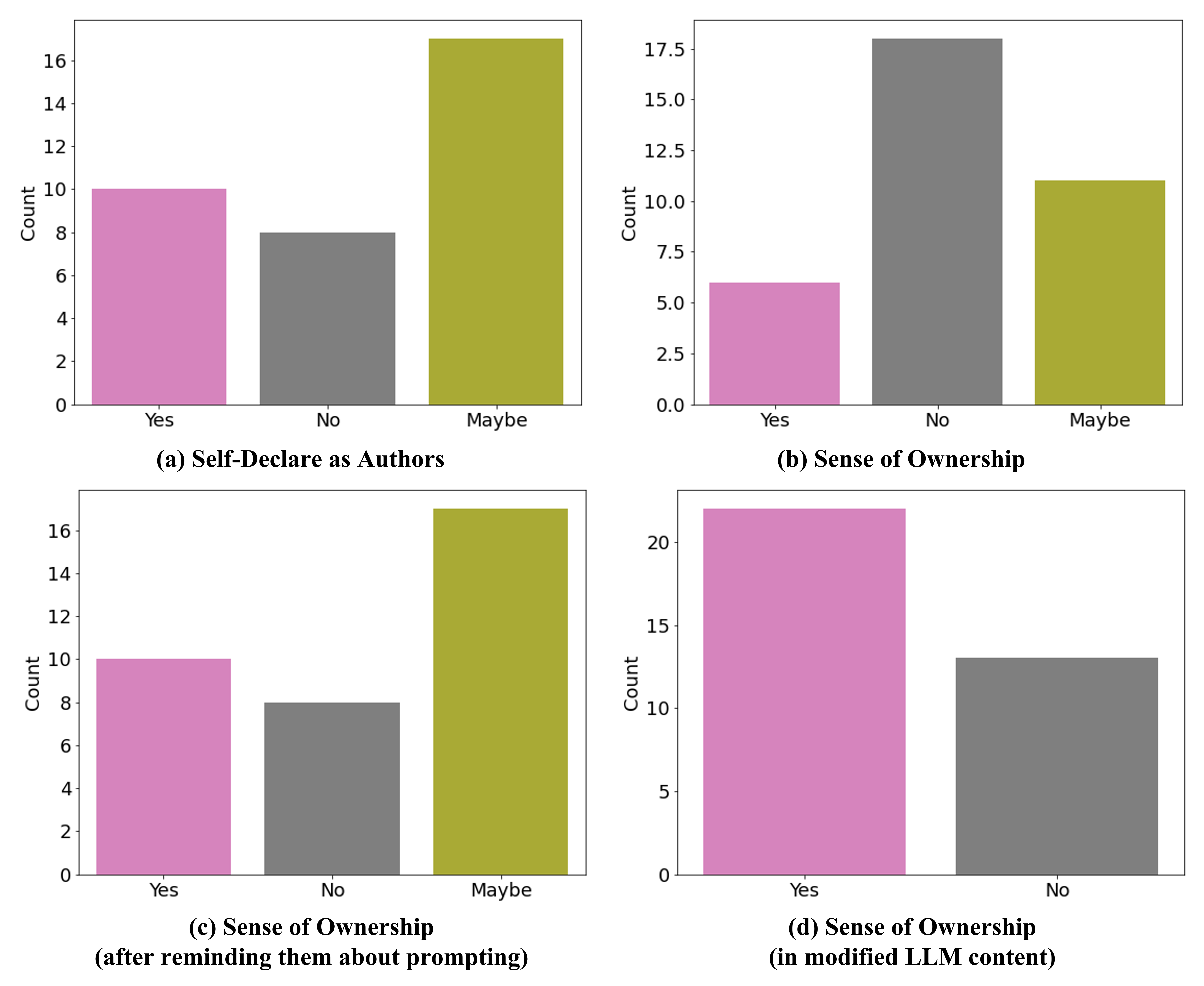}}
\caption{Survey Statistics: Different Levels of \textit{Sense of Ownership}}
\label{fig:11_12_13_14}
\end{figure}

Figure \ref{fig:11_12_13_14} (a,b) illustrates a compelling phenomenon: while individuals may not assert ownership of LLM content (\textit{'No':} 51.5\%, \textit{'Maybe':} 31.4\%), they demonstrate a willingness to submit identical LLM-generated content under their own authorship (\textit{'Yes':} 28.6\%, \textit{'Maybe':} 48.6\%). It implies that although individuals are willing to declare themselves as authors, they don't claim ownership of the content created by LLMs. The interesting psychological conflict of not claiming ownership while allowing to use their name requires further investigation, providing insight into complex notions of authorship and identity. 

\citet{10.1145/3637875}  also acknowledged and investigated this phenomenon and named it \textit{AI Ghostwriter Effect}, potentially influencing behaviors related to ghostwriting\footnote{
Ghostwriting is the practice of writing content on behalf of someone else who takes credit for the work, often without any acknowledgment of the ghostwriter's involvement.}, a common thing in writing industry \cite{doi:10.1177/107780040000600406}. This shows that there is a gap between acknowledging ownership and attributing authorship, which may have its roots in a desire for self-recognition or in the belief that authorship involves more than just possession—such as presentation or accountability. Though this is connected to our second research question, but our work distinctly focuses on disentangling the perceptions of ownership and authorship when using LLM writing assistants, an interesting aspect not directly explored in previous work. Our methodology is designed to explicitly probe this ownership-authorship dichotomy through carefully constructed experiments and surveys. This targeted approach allows us to deepen our understanding of human-centric aspects of LLMs.

After reminding participants of their role in providing the prompts (as discussed in Section \ref{sec: sur-q}), a significant shift occurred: more participants claimed ownership of the content ('Yes': increased from 17.1\% to 28.6\%; 'Maybe': increased from 31.4\% to 48.6\%; 'No': reduced from 51.4\% to 22.9\%, as shown in Figure \ref{fig:11_12_13_14} (c) ). So, reminding participants of their role in providing prompts likely increased their sense of ownership by highlighting their contribution to the content creation process, leading to a reevaluation of their ownership perception.
After reminding them about editing the LLM response, the sense of authorship surged, with 62.9\% of participants claiming ownership of the content (Figure \ref{fig:11_12_13_14} (d)).

The increase in participants claiming ownership after being reminded of their role in providing prompts suggests that highlighting their contribution to the content creation process enhances their sense of ownership. This reminder likely prompts individuals to reevaluate their ownership perception, recognizing their active involvement in shaping the content's direction. Similarly, the surge in the sense of ownership after being reminded about editing the LLM response indicates that individuals attribute ownership when they actively engage in refining and shaping the final output, reinforcing their sense of authorship and accountability. These reminders serve to emphasize the collaborative nature of content creation, influencing individuals' perceptions of ownership and authorship.


\section{Discussion}
The findings reveal a curious mismatch between owning something and being considered its author, hinting at some intriguing reasons behind it. It seems that people might feel conflicted because they want to be recognized for their work but are unsure if they should claim full ownership. This confusion might stem from their idea of what it means to be an author, which includes more than just saying "it's mine." However, when reminded of their role in creating and refining content, like suggesting ideas or editing responses, they start to feel more like owners and authors. This suggests that recognizing everyone's input is key to making people feel like they truly own and have authored something. Understanding this could help us create better ways for people to collaborate and feel proud of their work. Extensive use of LLMs also poses risks like opaque authorship, hallucination and bias introduction, inexplicability, loss of autonomy, critical reasoning, and thought outsourcing \cite{Bekker2024ww}.

Our goal is to lay the groundwork for better writing assistance systems and improve our understanding of human-centered aspects of writing. To achieve this, we need to increase transparency and accountability in content created with LLMs. Clear guidelines for disclosing LLM involvement can help people make informed decisions about ownership. Emphasizing collaborative approaches that blend human and AI contributions in creative tasks can strengthen feelings of ownership and authorship among individuals involved.

\textbf{Limitations.}
One limitation of this study is the small sample size, consisting of only 35 participants. Increasing the number of participants could lead to more robust and meaningful research outcomes. We also acknowledge a gap in content types, as only two types (general and creative) are explored. Future work in these areas should extend the content types and LLM application areas.

\section{Ethics and Laws}
Sense of ownership is intricately linked to tangible ownership, ethical considerations, and consequently intersects with legal frameworks and jurisdictions. And, this issue regarding AI-generated content remains complex and varies across jurisdictions. In the US, current law typically requires human authorship for ownership, as seen in the Thaler v. Perlmutter case. However, a Chinese court has ruled differently, granting AI-generated content protection under copyright law. Legal decisions and ongoing discussions shape our understanding of authorship and ownership in the AI era. Courts would likely examine how LLMs learn and produce content, potentially concluding that AI, not humans, authored the output, thus impacting ownership rights \cite{10.1145/3637875, Tombekai2020}.
For instance, OpenAI retains ownership of the language model and the content generated by ChatGPT, while users are acknowledged as collaborators but not the owners or authors of the AI-generated text \cite{10.1145/3637875}. 
This legal ambiguity may influence individuals' actual sense of ownership over AI-generated content as they grapple with the blurred lines between human input and machine output, potentially challenging traditional notions of ownership in the digital age.

\section{Findings and Conclusion}
This study explores into two prevalent mental dilemmas surrounding LLM-assisted writing: firstly, the discrepancy in crediting LLMs across various types of content, influencing individuals' sense of ownership; and secondly, the interplay between authorship claims and the perceived ownership of LLM-generated content. 
Below is a brief overview of conclusions supported by data regarding participants' motives, alongside some speculative insights:
\begin{itemize}
    \item Our research indicates that ownership perceptions vary based on content type when using LLMs as writing assistants. In creative works such as poems and stories, individuals tend to credit LLMs and feel less ownership. Conversely, in general content creation, individuals exhibit a stronger sense of ownership over the produced content.
    \item Our research reveals a interesting relationship between ownership and authorship perceptions when using LLMs as writing assistants. Individuals may not assert ownership over LLM-generated content but show a willingness to claim authorship.
    \item The experiments demonstrated a progression in ownership perception: participants were initially hesitant but willing to claim authorship. Reminders of their role in providing prompts increased ownership perception, and further engagement through editing the LLM response notably enhanced authorship perception, illustrating the impact of active involvement on ownership.
\end{itemize}
Through survey data analysis, in this work, we aim to identify prevalent dilemmas and underlying thought processes in human-centric aspects of LLM usage to advance understanding and enhance writing assistance system development.

\section*{Author Contributions}
ATW conceived the core idea, developed the methodology, designed the experiments and the data collection, performed some experiments, conducted all the analyses, prepared visualizations, wrote and edited core parts of the paper, and led the whole project. 
MRI collected the data, formal analysis and was involved in initial draft writing and editing.
RI helped in background study, writing and editing.

%

\bibliographystyle{ACM-Reference-Format}
\bibliography{our_work}

\appendix

\section{Demographic Profile of Survey Participants}  \label{sec: demographics}
The demographic profile of survey participants predominantly comprises university students aged 18-24 (94.3\%), with a smaller representation (5.7\%) in the 24-30 age group. The gender distribution skews towards males (65.7\%), while females constitute 34.3\% of the respondents. Notably, the vast majority (88.6\%) of participants come from a Science background, with a minority (8.6\%) from Arts backgrounds. It is worth mentioning that most individuals in this demographic rated their level of technology awareness as high, scoring between 3 and 5 on a scale where 5 represents a "tech nerd" and 1 indicates a lack of awareness of technology. This indicates a cohort of tech-savvy individuals, likely pursuing STEM fields in their university studies.

\end{document}